 \documentclass[aps,prb,twocolumn,showpacs]{revtex4-1}
\usepackage{graphicx}
\usepackage{dcolumn}
\usepackage{bm}
\usepackage{amsmath}

\newcommand{\e}{\varepsilon}

\usepackage{color}
\usepackage{float}
\usepackage{subfigure}
\usepackage{dsfont}
\usepackage{txfonts}
\usepackage{wasysym}
\usepackage{ifsym}
\usepackage{multirow}
\newcommand{\vernek}[1]{\textcolor{black}{#1}}

\begin{document}

\title{Parity oscillations of Kondo temperature in a single molecule break junction}
\author{B.~M.~F.~Resende}
\address{Instituto de F\'i­sica - Universidade Federal de Uberl\^andia -
Uberl\^andia, MG  38400-902  - Brazil}
\author{E.~Vernek}
\email[Corresponding author:]{vernek@infis.ufu.br}
\address{Instituto de F\'i­sica - Universidade Federal de Uberl\^andia -
Uberl\^andia, MG  38400-902  - Brazil}

\date{\today}

\begin{abstract}
We study the Kondo temperature ($T_K$) of  a single molecule break junction. By employing a
numerical
renormalization group calculations we have found that $T_K$ depends
dramatically upon the position of the molecule in the wire formed between the contacts. We
show that $T_K$ exhibits strong \emph{oscillations} when the parity of the left {and/or} right
number of atomic sites ($N_L,N_R$)  is changed. For a given set of parameters, the maximum value of
$T_K$ occurs for ($odd,odd$) combination, while its minimum values is observed for ($even,even$).
 These oscillations are fully understood in terms of the effective hybridization function. 
\end{abstract}

\maketitle

Kondo effect (KE) is one of the most intriguing phenomena of
strong correlated systems,\cite{Hewson-Kondo} which was beautifully explained  by J. Kondo in the
60's in the seminal theoretical work on the minimal resistance in
magnetic alloys.\cite{Prog.Theor.Phys...32} KE has revived in the later 90's  with the
advent of the scanning tunneling microscope (STM) that has facilitated the manipulation of the
matter at atomic scale. For instance, STM has allowed observation of interesting facets of the  KE
in quantum dots (QD)\cite{Nature.391.156,*Science.S.1998.540} and in single atom or molecule on
metallic surfaces,\cite{Sicence.280.567,Nature.403.512,PhysRevLett.97.266603} which have motivated a
huge number of experimental\cite{Nature.391.156,Science.S.1998.540,Sicence.268.1440,
*Science.293.2221} and theoretical\cite{PhysRevLett.82.3508,PhysRevB.78.054445}
investigations. 

One of the experimentally accessible  signatures of the KE in nanoscopic system  
 such as QD or magnetic molecules attached to metallic contacts is the
strong modification in  the conductance across the system, observable when the system is cooled
down below the {so-called} Kondo temperature ($T_K$). In QD, for instance, $T_K$ is found to be in
the sub Kelvin region whereas for large molecules attached to metallic leads
$T_K$ can be much larger.\cite{PhysRevLett.97.266603} In both cases, in
the limit of very strong Coulomb interaction, $T_K$ depends
strongly upon the effective hybridization ($\Delta$) that connects the localized magnetic moments
to the conduction electrons\cite{PhysRevLett.97.096603,PhysRevLett.97.266603}  as\cite{Hewson-Kondo}
$T_K\sim \exp({\pi\varepsilon_d/\Delta})$, where $\varepsilon_d$ ($<0$) is the energy of the
localized orbital respect to the Fermi level. 
Controlling $\Delta$ or $\varepsilon_d$ is therefore crucial for obtaining higher $T_K$, which is
fundamental for possible  technological application of  KE. While tuning
$\e_d$ is relatively simple in QDs by mean of gate voltages, in molecules, on the other
hand, it becomes a more complicated task. Conversely, geometrical parameters are
more suitably modified in molecules than in QDs and  has proven to produce important
modifications in $T_K$ {via} hybridization function.\cite{PhysRevLett.99.026601} 

\begin{figure}[!htb]
\vskip0.1cm
\centerline{\resizebox{3.in}{!}{
\includegraphics{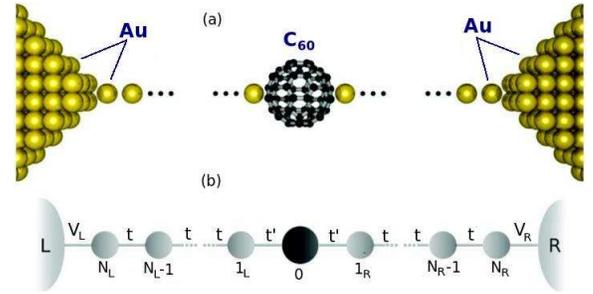}}}
\caption{\label{sistema}(Color online) (a) Illustration of a Au quantum
wire coupled to metallic contacts with a embedded C$_{60}$ molecule. (b) Pictorial
representation of the model. The site labeled as ``0'' represents the impurity site with strong
on-site coulomb repulsion.} 
\end{figure}

A suitable
experimental arrangement to study the KE is the break junction (BJ) molecular structures, in which a
metallic wire (gold wire, for instance) is stretched until a few-atom 1D chain bridges the gap
between the electrodes before the complete break up of
the wire.\cite{PhysRevLett.87.096803,NatureCommunications.2.305,NatureNanotechnology.1.173,
Nature.419.906,PhysRevLett.91.076805} Owing to the dependence of $T_K$ upon $\Delta$ it has been
shown that $T_K$ can be mechanically modulated in BJ experiments\cite{PhysRevLett.99.026601} by
changing the distance between the electrodes. Motivated by this experiment, in the present
work we study the Kondo temperature of a spin-$1/2$ magnetic impurity coupled to metallic
contacts through two finite (left and right) quantum wires (QW), as illustrated in
Fig.~\ref{sistema}(a). By employing a numerical
renormalization
group\cite{RevModPhys.47.773,RevModPhys.80.395} (NRG) calculation we find a strong dependence of
$T_K$ 
upon the parity of the number of sites
($N_R,N_L$) of each QW as well as their length. The dependence upon the ($N_L,N_R$)
parity combination results in an oscillating behavior of $T_K$ as function of $N_L$ or $N_R$, akin
to what has been observed in Manganese phthalocyanine (MnPc) molecules on top of Pb islands,
reported in Ref.~\onlinecite{PhysRevLett.99.256601}. 
Although the system under investigation here is quite different from the one studied in
Ref.~\onlinecite{PhysRevLett.99.256601}, the origin of the oscillation of $T_K$ can be
interpreted likewise. While in their case the enhancement of $\Delta$ originates from the
formation of multiple quantum well states between the Pb atomic layers, here the enhancement of
$\Delta$
results from the localized states of the atomic sites of the QW. 

%
Our system model is schematically represented in Fig.~\ref{sistema}(b) and is modeled by the
Anderson-type Hamiltonian that can be split into five terms  as 
\begin{eqnarray}
H=H_{imp}+H_{cb}+H_{wires}+H_{imp-wires} +H_{cbs-wires}, 
\end{eqnarray}
 where $H_{imp}$, $H_{cb}$ and $H_{wires}$ describe, respectively, the interacting impurity,  
the free electrons in the conduction bands and the electrons in the wires, $H_{imp-wires}$ couples
the impurity to the two wires and $H_{cb-wires}$ couples the wires the their respective conduction
bands. In terms of creation and annihilation fermion operators the Hamiltonians read
\begin{eqnarray}\label{H_imp}
 H_{imp}&=&\sum_{\sigma}\varepsilon_dc^\dagger_{d\sigma}c_{d\sigma}+Un_{
d\uparrow}n_{
d\downarrow},\\
H_{cb}&=&\sum_{\ell=R,L}\sum_{k\sigma}\varepsilon_kc^\dagger_{\ell k\sigma}c_{\ell
k\sigma}\\
H_{wires}&=&
\sum_{\ell=R,L}\left[\vernek{\varepsilon_0\sum_{i_\ell=1\atop\sigma}^{N_\ell}n_{i_\ell\sigma}+}
t\sum_{i_\ell=1\atop \sigma}^{N_\ell-1}\left(c^\dagger_{i_\ell\sigma}c_{
i_\ell+1} +H.c.\right)\right],\\
 H_{cb-wires}&=&\sum_{\ell=R,L}\sum_{\ell k\sigma} \left(V_{\ell k}c^\dagger_ { N_{\ell}
}c_
{\ell k\sigma}+V^*_{\ell k} c^\dagger_{
\ell k\sigma}c_{ N_\ell }\right)\\
\label{H_imp_wires}
H_{imp-wires}&=&t^\prime \sum_{\ell=R,L}\sum_{\ell
\sigma}\left(c^\dagger_{d\sigma}c_{1_\ell}+c^\dagger_{1_\ell\sigma}c_{
d\sigma } \right).
\end{eqnarray}
In Eqs.~\ref{H_imp}-\ref{H_imp_wires}, the operators $c^\dagger_{d\sigma}$ ($c_{d\sigma}$) creates
(annihilates) an electron in the orbital $d$ with energy $\varepsilon_d$,
$c^\dagger_{\ell k\sigma}$ ($c_{\ell k\sigma}$) creates (annihilates) an electron in the $\ell${th}
conduction band with energy $\varepsilon_k$, $c^\dagger_{i_\ell\sigma}$ ($c_{i_\ell\sigma}$)
creates (annihilates) and electron in the $i${th} site of the $\ell${th} QW with energy
$\varepsilon_0$ spin $\sigma$.
Finally, $t$ is the hopping between two
adjacent sites in the wires and $V_{\ell k}$ and $t^\prime$ couple the QWs to
their conduction bands and to the impurity, respectively. The conduction bands are
characterized by a flat density of states, $\rho(\omega)=(1/2D)\Theta(D-|\omega|)$, where
 $D$ is their half bandwidth and $\Theta(x)$ is the Heaviside step function.   
It is worth emphasizing that although the motivating experiment was realized using C$_{60}$ molecule
coupled to Au metallic contacts, this model is rather general. In the particular context of
molecular BJ, vibrations may  be important in certain range of parameters, but this aspect is beyond
the scope of the present work.

In order to properly address the Kondo physics of the system, the full Hamiltonian is
approached by using the numerical renormalization method, with which we can
calculate the thermodynamical properties. 
Within the NRG approach we  discretize the effective
conduction band ``seen'' by the interacting impurity. The effective conduction
band can be determined by exact calculation of the local non-interacting ($U=0$) Green's
function (suppressing the spin index), 
$g_{dd}(\omega)=[\omega-\epsilon_d+\Sigma(\omega)]^{-1}$, 
where $\Sigma=\Sigma_R(\omega)+\Sigma_L(\omega)$, in which the $\ell$th self-energy is given by
\begin{eqnarray}\label{Sigma}
\Sigma_\ell(\omega)=-\cfrac{t^{\prime 2}}{\omega-\cfrac{t^2}{
\omega-\cfrac { t^2}{ \omega-\cfrac{t^2}{ \cfrac{\ddots t^2}{ \omega-V_\ell^2\tilde
g(\omega)}}}}},
\end{eqnarray}
with
\begin{eqnarray}
\tilde g(\omega)=-\frac{1}{2D}\ln\Big|\frac{\omega-D}{\omega+D}\Big|
-\frac{i\pi}{2D}\Theta(D-|\omega|)
\end{eqnarray}
being the diagonal GF associated to the unperturbed conduction electrons in the
leads. The fraction in Eq.~\ref{Sigma} is continued until all the sites of the  of the $\ell$th
wire and the $\ell$th conduction band are taken into account. 

The hybridization of the localized orbital ``$d$'' with the effective band is
give by $\Delta(\omega)={\tt
Im}[\Sigma(\omega)]=\Delta_L(\omega)+\Delta_R(\omega)$ [Hereafter we will refers to $\Delta(0)$
just as $\Delta$].
The hybridization function is logarithmically
discretized\cite{RevModPhys.80.395,PhysRevB.52.14436} to map the system in a
Wilson's chain form,\cite{RevModPhys.47.773}
\begin{eqnarray}
 H=H_{imp}+t^\prime\sum_{\sigma}\left(c^\dagger_{d\sigma}c_{0\sigma}+c^\dagger_{0\sigma}c_{d\sigma}
\right)+\sum_{i=0\atop \sigma }^\infty\varepsilon_{i\sigma}c^\dagger_{i\sigma}c_{i\sigma}
\nonumber\\
+\sum_{i=0\atop \sigma }^\infty
t_i\left(c^\dagger_{i\sigma}c_{i+1\sigma}+c^\dagger_{i+1\sigma}c_{i\sigma}\right),
\end{eqnarray}
where $t_i$'s are calculated via $\Delta(\omega)$, following the recipes described in
Ref.~\onlinecite{PhysRevB.52.14436}.  Once we have mapped the system in the Wilson's form, we
proceed the
NRG calculation, which is based in the iterative diagonalization of the effective
Hamiltonian.\cite{RevModPhys.80.395} After reaching the strong coupling fixed point we can calculate
the magnetic moment within the canonical ensemble as
\begin{eqnarray}\label{mmoment}
 k_BT\chi(T)=\frac{1}{Z(T)}\sum_{\nu}\left[\langle
\nu|S^2_z|\nu\rangle-(\nu|S_z|\nu\rangle)^2\right]
e^{-E_\nu/k_BT},
\end{eqnarray}
where $k_B$ is the Boltzmann constant,  $Z(T)=\sum_\nu\exp(-E_\nu/k_BT)$ is the canonical partition
function, $S_z$ is the spin
operator, $|\nu\rangle$ and $E_\nu$ are, respectively, the eigenvector and its corresponding
eigenvalue of the full interacting Hamiltonian, which are naturally calculated in the NRG
procedure.\footnote{All the results were obtained using the conventional NRG
 discretization parameter $\Lambda=2.5$ and keeping typically $1500$ states at each iteration.}
Following Wilson's criterion, we define
$T_K$ from the  ``impurity'' magnetic moment as $k_BT_K\chi_{imp}(T_K)=0.0707(g\mu_B)^2$, (that is
the magnetic moment of the full system subtracted by the contribution of
the effective conduction band), $g$ is the electron $g$-factor and $\mu_B$ is the Bohr magneton.

Before starting the presentation of our numerical results, lets analyze the hybridization function
at the Fermi level, which is the most  relevant parameter to determine the behavior of $T_K$ in our
calculations.  It is straightforward to show from Eq.~\ref{Sigma} that $\Delta$ possesses only
three distinct vales, 
\begin{equation}
 \Delta=\left\{
\begin{array}{cc}
\Delta_{min}=\Delta_0  &\quad \mbox{for (even,even)}\\
\Delta_{int}=\Delta_0\left(\frac{1}{2}+ \alpha\right) & \quad \mbox{for (even,odd) or
(odd,even)}\\
\Delta_{max}=2\alpha\Delta_0 & \quad \mbox{for (odd,odd)}
\end{array}
\right.,
\end{equation}
where we have defined $\Delta_0=\pi t^{\prime 2}/D$ and denoted $\Delta_{min}$, $\Delta_{int}$ and
$\Delta_{max}$, the minimum, intermediate and maximum
value of $\Delta$, respectively, and  $\alpha=2(D/\pi t)^2$ is a dimensionless
parameter that can be modified, for instance, by stretching the QW as is was done in the
Ref.~\onlinecite{PhysRevLett.99.026601}.  To obtain our numerical results lets set $D=1$ as our
energy scale. With that we choose hereafter  \vernek{(unless otherwise stated)} $U=0.5$,
$\epsilon_{d}=-0.25$, $\varepsilon_0=0$ (at the particle-hole symmetric
point), $V_R=V_L=t=0.15$ and $t^\prime=0.1$. 

\begin{figure}
\vskip0.1cm 
\centerline{\resizebox{3.in}{!}{
\includegraphics{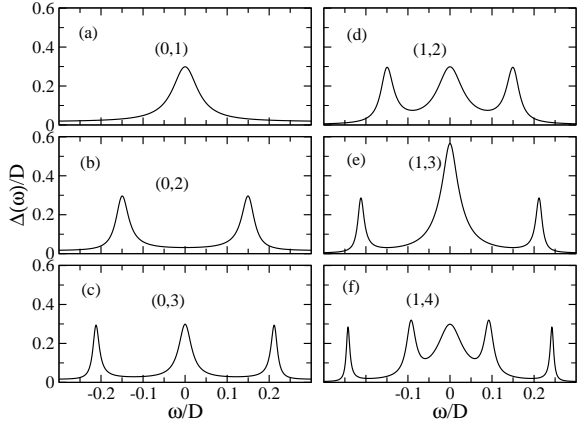}}}
\caption{\label{fig2}(Color online) Hybridization function vs energy for various values of
$N_{L}$ and $N_L$ [denoted in the figure as ($N_L,N_R$) using $V_{L}=V_{R}=t=0.15$ and
$t^\prime=0.1$. Notice that
$\Delta$ possesses three different values, depending on the parity of $N_L$ and $N_R$. The
minimum ($\Delta_{min}$) and maximum ($\Delta_{max}$) value of $\Delta$ is obtained for
$(even,even)$ (b) and  $(odd,odd)$ (e), respectively, while for all the other combinations $\Delta$
has an intermediate value, $\Delta_{int}$.}
\end{figure}

In Fig.~\ref{fig2} we show the hybridization function vs energy for various values of $N_L$ and
$N_R$. In Figs.~\ref{fig2}(a), \ref{fig2}(b), and \ref{fig2}(c)  we fix $N_L=0$ and
show $\Delta(\omega)$ for $N_R=1$, $N_R=2$ and $N_R=3$, while in Figs.~\ref{fig2}(d), \ref{fig2}(e),
and \ref{fig2}(f) we keep $N_L=1$ fixed  and show $\Delta(\omega)$ 
 for $N_R=2$, $N_R=3$ and $N_R=4$. The number of peaks of $\Delta(\omega)$ is given by
$\max(N_L,N_R)$ for equal parity and $N_R+N_L$ for different parities. Although the structure of
$\Delta(\omega)$ away from the Fermi level has some effect on $T_K$, the most relevant contribution
comes from the structures \emph{at} or \emph{very close} the the Fermi level.
For the parameters set above, we obtain $\Delta_{min}\approx 0.0314$, $\Delta_{max}\approx 0.566$
and $\Delta_{int}\approx 0.299$. These distinct values of
$\Delta$ are crucial for determining the Kondo temperature of the system, which in our case can be
roughly estimated as\cite{PhysRevB.21.1003} $k_BT_K(\Delta)\sim\exp[-\pi U/8\Delta]$. It is clear
that
$T_K$ increases as $\Delta$ increases.
\begin{figure}
\vskip0.1cm
\centerline{\resizebox{3.in}{!}{
\includegraphics{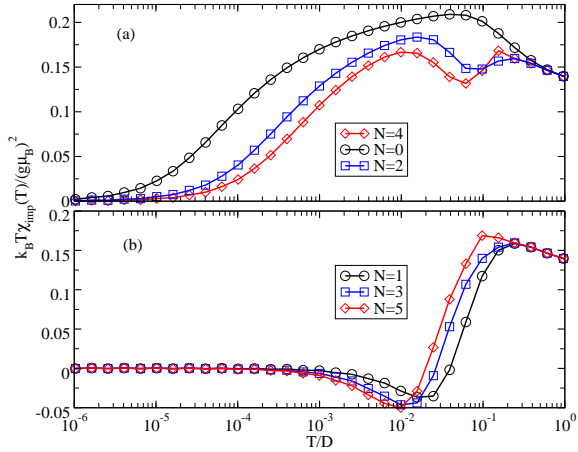}}}
\caption{\label{moment}(Color online) Magnetic moment as function of temperature for various
values of $N_R=N_L=N$. Panel (a) and (b) correspond to $N$ even  and odd,
respectively (see values of $N$ in the legends).}
\end{figure}

In Fig.~\ref{moment}(a) and Fig.~\ref{moment}(b) we show the magnetic moment as function temperature
for various values of $N=N_L=N_R$ (the symmetric case) even and odd, respectively. The case of
$N=0$ [Fig.~\ref{moment}(a), $\Circle$ (black)] corresponds to the single impurity coupled to two
conduction band. The low temperature suppression in the magnetic moment results from the Kondo
screening of the local spin [these curves are used to $T_K$, as discussed \vernek{above}].
On the other hand, in the high temperature limit the
$k_BT\chi\rightarrow (g\mu_B)^2/8$, as expected. Notice in Fig.~\ref{moment}(a)
that $T_K$ increases when $N$ (even)
increases. Conversely, $T_K$ decreases when $N$ (odd) increases as seen in Fig.~\ref{moment}(b).
Notice also that $T_K$ can be at least two order of magnitude larger for $N$ odd than for $N$ even.
This huge difference will be analyze below. The
result for  $N=1$ [Fig.~\ref{moment}(b), $\Circle$ (black)] is equivalent to those reported in
Ref.~\onlinecite{PhysRevLett.97.096603}. The negative values of $k_BT\chi_{imp}$ within a small
range of $T$ results from the subtraction of the effective conduction band contribution.

\begin{figure}
\vskip0.1cm
\centerline{\resizebox{3.in}{!}{
\includegraphics{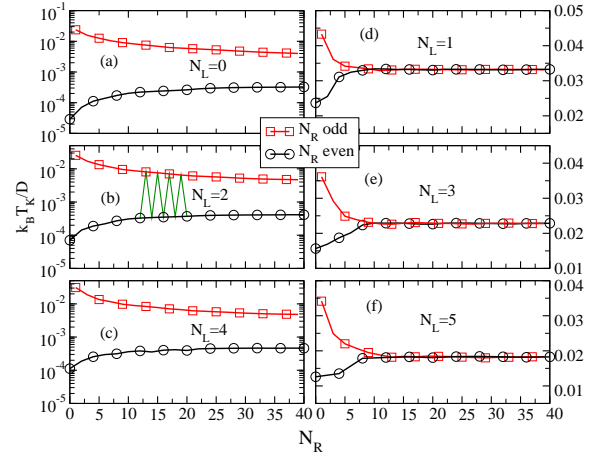}}}
\caption{\label{fig3}(Color online) Kondo temperature as function of
number of sites ($N_{R}$) for a fixed $N_{L}=1$. $\Circle$ (black) and $\square$ (red)
correspond to even (odd) $N_R$, respectively. The zig-zag (green) line shows the ($even,odd$)
oscillation, similar to those observed in Ref.~\onlinecite{PhysRevLett.99.256601}.}
\end{figure}

\begin{figure}
\vskip0.1cm
\centerline{\resizebox{3.0in}{!}{
\includegraphics{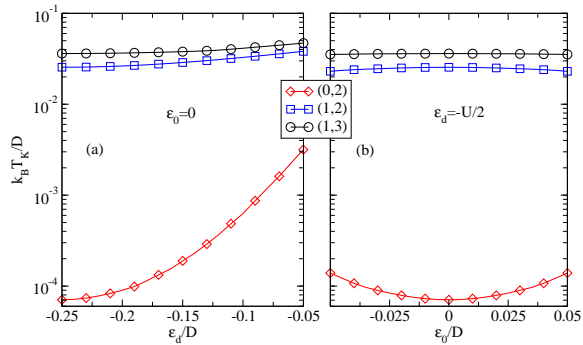}}}
\caption{\label{fig5}(Color online) Kondo temperature as function of
$\varepsilon_d$ ($\varepsilon_0=0$) (a) and $\varepsilon_0$ ($\varepsilon_d=-U/2$) (b) for
various configuration of $(N_L,N_R)$ as shown in the legend. The other parameter are the same as in
the previous figures. }
\end{figure}

In order to show the  behavior of the Kondo temperature for larger and different
values of $N_L$ and $N_R$ we show in Fig.~\ref{fig3} $T_K$ as function of $N_R$ for a fixed number
$N_L$ even (left) and $N_L$ odd (right). The $\Circle$ (black) curves correspond to $N_R$ even,
while $\square$ (red) curves corresponds to $N_R$ odd. Notice that for $N_L$ even
 [(Fig.~\ref{fig3}(a), \ref{fig3}(b) and \ref{fig3}(c)] $T_K$ increases with $N_R$ even
[$\Circle$ (black)]
while it decreases for  $N_R$ odd [$\square$ (red)]. Observe again that for small
$N_R$ $T_K$ is almost two order of magnitude larger for $N_R$ odd than for $N_R$ even (keeping $N_L$
even). This difference decreases asymptotically  for large $N_R$ and vanishes asymptotically as
$N\rightarrow\infty$. 
This results from the fact that in this situation the conduction electrons near the Fermi level are
more strongly coupled to the impurity, reflecting the fact that $\Delta_{int}$ is larger than
$\Delta_{min}$ as clearly shown in Fig.~\ref{fig2}. For $N_L$ odd (Fig.~\ref{fig2}(d),
\ref{fig2}(e) and \ref{fig2}(f) we observe a similar behavior ($T_K$ increases as $N_R$ even
increases  and decreases as
$N_R$ even increases) but in this case the curves collapse onto each other very quickly (typically
for $N_R=10$) to a larger value, when compared to the case of $N_L$ even. At least  for
small $N_L$ and $N_R$ we can roughly estimate the ratio between $T_K$'s for the three distinct
values of $\Delta$ as
\begin{eqnarray}
 \frac{T_K(\Delta_{a})}{T_K(\Delta_{b})}=e^{\frac{\pi
U}{8}\left(\frac{\Delta_a-\Delta_b}{\Delta_a\Delta_b}
\right)},
\end{eqnarray}
where $a$ and $b$ stand for $min$, $int$ and $max$. Using the parameters chosen above we
obtain
$T_K(\Delta_{int})/T_K(\Delta_{min})\approx 2.7\times10^2$,
while $T_K(\Delta_{max})/T_K(\Delta_{int})\approx 1.36$. These values are consistent with the
huge difference
between the values shown in $\square$ (red) and $\Circle$ (black) curves of Figs.~\ref{fig3}(a),
\ref{fig3}(b) and \ref{fig3}(c)  and small difference in the related curves of Figs.~\ref{fig3}(d),
\ref{fig3}(e) and \ref{fig3}(f). 
The behavior of $T_K$ with increasing $N_L$ and $N_R$ for the same parity combination
cannot be explained in terms of $\Delta$. This can be reasonably understood in terms of 
the formation of a small sub-band inside the conduction band,  due to a large number of atomic
sites and the energy dependence of hybridization function near the Fermi level. In the limit of
$N_L,N_R\rightarrow \infty$ the sub-band becomes a smooth curve, resulting in a
$T_K$ independent of the  lengths of the wires. The zig-zag (green) line in
Fig.~\ref{fig3}(b) shows the even-odd oscillations in $T_K$, very similar to what was observed in
Ref.~\onlinecite{PhysRevLett.99.256601}. \vernek{Finally, in Fig.~\ref{fig5} we show the robustness
of these results against particle-hole symmetry breaking. In Fig.~\ref{fig5}(a) show $T_K$ as
function of $\varepsilon_d$ for $\varepsilon_0=0$. Notice that, although more pronounced for the
$(even,even)$ case, $T_K$ increases as $\varepsilon_d$ is shifted upward from $-U/2$ for all
parities
[(0,2), (1,2) and (1,3)]. Same behavior is obtained for the other side (not shown). These are
consistent with the general expression,\cite{PhysRevLett.40.416} $T_K\sim
Exp{\left[-\pi|\varepsilon_d|(\varepsilon_d+U)/(2\Delta U)\right]}$ for constant hybridization
function. When we keep $\varepsilon_d$ and vary $\varepsilon_0$ about the Fermi level
[Fig.~\ref{fig5}(b)] we see that $T_K$ increases for $(0,2)$ but decrease slightly for $(1,2)$ and
$(1,3)$. This results from the fact that, as $\varepsilon_0$ deviates
from the Fermi level, $\Delta(0)$ decreases if $\Delta(\omega)$ possesses a peak at the Fermi
level as in the $(even,odd)$, $(odd,even)$ or $(odd,odd)$ cases, but increases  when
$\Delta(\omega)$ exhibits a dip at the Fermi level as in the $(even,even)$ configuration.
}

%


\vernek{In conclusion}, we have presented a detailed study of the Kondo temperature of a single
molecule
break junction. By employing a numerical renormalization group we show that $T_K$ is strongly
dependent of the parity of the number of atomic sites in each piece of QW connecting the molecule to
the contacts. More interesting, we show that the $T_K$ oscillates when the parity of the number of
sites of the wires changes. These oscillations are  interpreted in terms of the effective
hybridization function $\Delta(\omega)$. For $(even,even)$ and $(odd,odd)$ configurations the
effective coupling $\Delta$ is minimum and maximum, respectively, while for $(even,odd)$ or
$(odd,even)$ configurations $\Delta$ possesses an intermediate value.  Within this picture,
the huge variation of $T_K$ is  readily estimated by a simple analytical calculation, which
can vary up to a factor of $10^2$ [in the case of changing from ($even,even$) to ($even,odd$)]. Our
results provide a very clear picture of the main ingredient responsible for the dramatic
dependence of $T_K$ on geometrical configuration of single molecule break junctions as well as of
magnetic molecule on atomic layer surfaces. Moreover, we believe our results can be
used to guide experimental realizations of high-$T_K$ experiments.

We would like to thank  CNPq (under grant No. 493299/2010-3) and FAPEMIG (under grant No.
CEX-APQ-02371-10) for financial support. We also  wish to acknowledge valuable discussions with F.
M. Souza.

%

\end{document}